\documentclass[journal]{IEEEtran}
\usepackage{algorithm}
\usepackage{algorithmicx}
\usepackage{algpseudocode}
\usepackage{amsmath}
\usepackage{mathrsfs}
\usepackage{amsfonts}
\usepackage[numbers,sort&compress]{natbib}
\usepackage{amssymb}
\usepackage{bm}
\usepackage{enumerate}
\usepackage[pdftex]{graphicx}
\usepackage{lineno}
\usepackage{url}
\emergencystretch=\maxdimen
\ifCLASSINFOpdf
\else
\fi

\begin{document}
%
\title{Virtual Codec Supervised Re-Sampling Network for Image Compression}

%

\author{Lijun~Zhao,
        Huihui~Bai,~\IEEEmembership{Member,~IEEE,}
        Anhong~Wang,~\IEEEmembership{Member,~IEEE,}
        and~Yao~Zhao,~\IEEEmembership{Senior~Member,~IEEE}

\thanks{Corresponding author: Huihui Bai. This work was supported in part by National Natural Science Foundation of China (No. 61672087, 61672373), Key Innovation Team of Shanxi 1331 Project （ KITSX1331） and the Fundamental Research Funds for the Central Universities (No. 2017YJS053).}
\thanks{L.  Zhao, H. Bai, Y. Zhao are with the Beijing Key Laboratory of Advanced Information Science and Network Technology, Institute Information Science, Beijing Jiaotong University, Beijing, 100044, P. R. China, e-mail: {15112084, hhbai, yzhao}@bjtu.edu.cn.}
\thanks{A. Wang is with Institute of Digital Media \& Communication, Taiyuan University of Science and Technology, Taiyuan, 030024, P. R. China, e-mail: wah\_ty@163.com}
}
%
%

\markboth{Journal of \LaTeX\ Class Files}
{Shell \MakeLowercase{\textit{et al.}}: Bare Demo of IEEEtran.cls for IEEE Journals}
%



\maketitle

\begin{abstract}
In this paper, we propose an image re-sampling compression method by learning virtual codec network (VCN) to resolve the non-differentiable problem of quantization function for image compression. Here, the image re-sampling not only refers to image full-resolution re-sampling but also low-resolution re-sampling. We generalize this method for standard-compliant image compression (SCIC) framework and deep neural networks based compression (DNNC) framework. Specifically, an input image is measured by re-sampling network (RSN) network to get re-sampled vectors. Then, these vectors are directly quantized in the feature space in SCIC, or discrete cosine transform coefficients of these vectors are quantized to further improve coding efficiency in DNNC. At the encoder, the quantized vectors or coefficients are losslessly compressed by arithmetic coding. At the receiver, the decoded vectors are utilized to restore input image by image decoder network (IDN). In order to train RSN network and IDN network together in an end-to-end fashion, our VCN network intimates projection from the re-sampled vectors to the IDN-decoded image. As a result, gradients from IDN network to RSN network can be approximated by VCN network's gradient. Because dimension reduction can be further achieved by quantization in some dimensional space after image re-sampling within auto-encoder architecture, we can well initialize our networks from pre-trained auto-encoder networks. Through extensive experiments and analysis, it is verified that the proposed method has more effectiveness and versatility than many state-of-the-art approaches.
\end{abstract}


\IEEEpeerreviewmaketitle
\section{Introduction}

Advanced techniques about 3D video \cite{r1}, 360 panorama video \cite{r2}, light field \cite{r3}, etc., have received more and more attentions and have widely researched due to their practically applied values. However, the information carrier of these techniques mainly refers to image, thus Internet congestion may occur, because of explosive growth image data among social media and other new media. With this trend of rapidly increasing, the main source of Internet congestion will be caused by image/video transmission \cite{r4}, so different kinds of images, especially natural image, should be extremely compressed to alleviate this problem.

Image compression aims at reducing the amounts of data to benefit image storage and transmission. Still image compression has been developed from early image compression standards such as JPEG and JPEG2000 to Google's WebP and BPG, etc. In the earlier times, a lot of works \cite{r5, r6, r7, r8, r9, r10, r11, r12, r13, r14, r15, r16, r17} mainly put their emphasis on post-processing to reduce coding artifacts so as to improve coding efficiency, whose priority exists in that it doesn't need change any part of existing coding standard. Lately, several works \cite{r18, r19, r20, r21, r22, r23, r24, r25} employ convolutional neural network (CNN) to remove image blurring and quantization artifacts caused by image compression. Among these works, a very special work \cite{r25} is an effective compression framework based on two collaborative convolutional neural networks, where one network is used to compactly represent image and the other one works as post-processing to reduce coding distortion. This method has good performances at the case of very low bit-rate coding, but it doesn't explore how to improve coding efficiency at high bit-rate. Thus, this method's practical application is very limited, because image coding at low bit-rate is required, only when the band-width is very narrow. Meanwhile, this method directly trains collaborative convolutional neural networks without considering quantization's effects on neural network ahead of standard codec during back-propagation, so it's a sub-optimal solution for image compression.

Recently, image compression with deep neural networks (DNN) has achieved many great breakthroughs, such as \cite{r27, r28, r29, r30, r31, r32, r33, r34}, among which some methods have exceeded JPEG2000 and even can compete with BPG. These methods target at resolving the challenging problem: quantization function within objective compression loss is non-differentiable. The pioneering work \cite{r27} leverages recurrent neural networks to compress image in full-resolution, where the binarization layer with stochastic binarization is used to back-propagate gradients. In \cite{r29, r30}, the quantizer in general nonlinear transform coding framework is replaced by an additive independent identically distributed uniform noise, which can make image compression optimized by gradient descent method. In \cite{r31}, identity smooth function's derivation is used as an approximation of rounding function's derivation in the compressive auto-encoders, but no modification is required during passing gradients from decoder to encoder. Most recently, soft assignment is formed by converting Euclidean distance between vector and each quantization center into probability model via soft-max function \cite{r32}. After that, soft quantization is defined by soft assignment, and then this smooth relaxation is used as the approximation of the quantized function, so that compression loss of auto-encoder networks in terms of quantization can be optimized by stochastic gradient descent method.

\begin{figure}[t]
\centering
\includegraphics[width=3in]{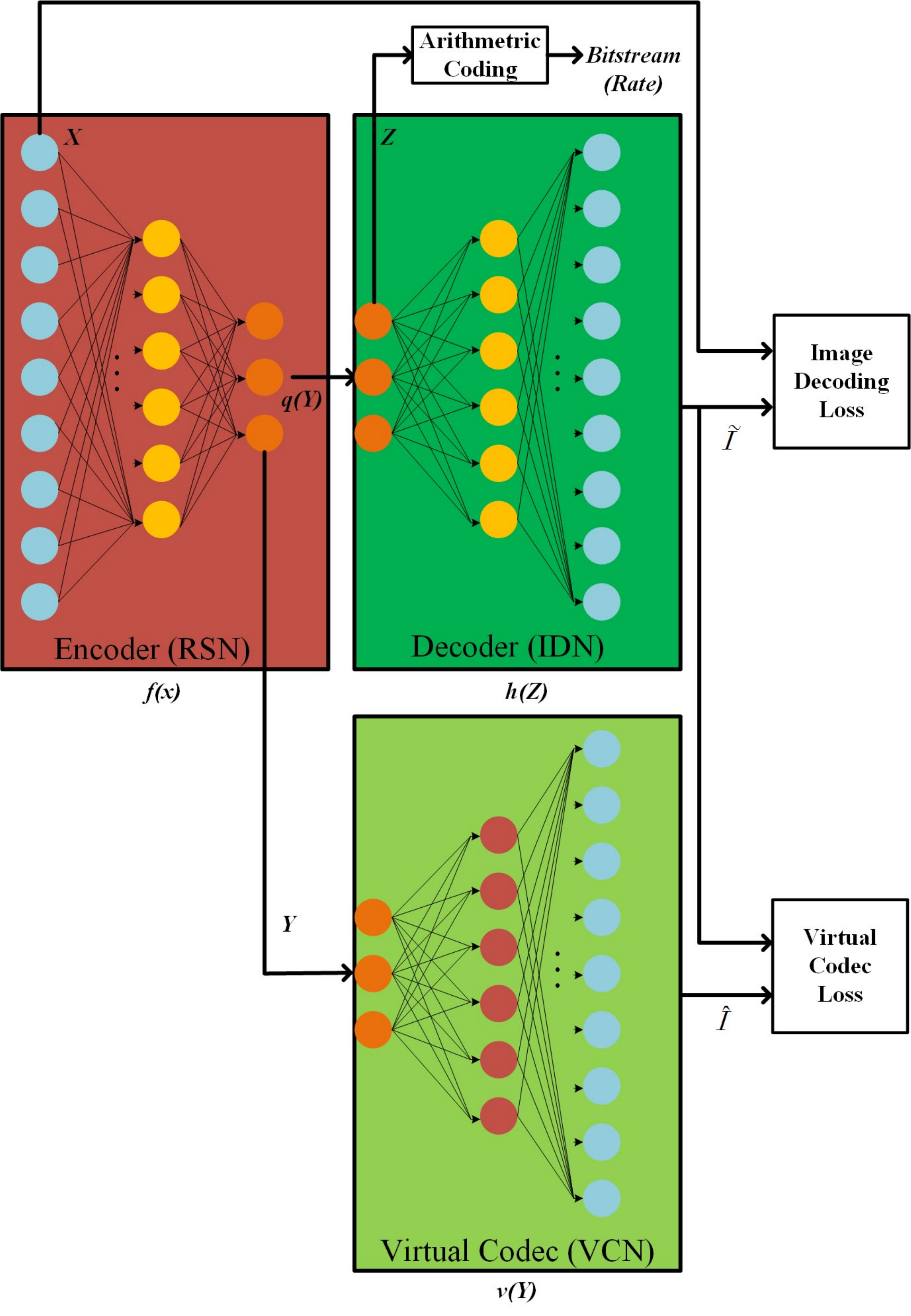}
\caption{The diagram of deep neural networks based compression framework}
\label{Fig1}
\end{figure}

Our intuitive idea is to learn projection from re-sampled vector to the quantized vector, so that we can jointly train our RSN network and IDN network together. However, we find it's difficult to learn this projection directly with DNN. Fortunately, the projection can be well intimated by neural network from the above re-sampled vectors to the decoded image. Therefore, we propose an image re-sampling compression method (IRSC) by learning virtual codec network (VCN) to supervise re-sampling network (RSN) to resolve the non-differentiable problem of quantization function within compression loss. For simplicity, we give a diagram of deep neural networks based compression framework (DNNC) for one dimension signal, as shown in Fig.\ref{Fig1}.

Our IRSC method can not only be used for DNNC framework, but it also can be applied to standard-compliant image compression framework (SCIC). Firstly, an input image is measured by RSN network to get re-sampled vectors. Secondly, these vectors are directly quantized in the re-sampling feature space for DNNC, or transformation coefficients of these vectors are quantized to further improve coding efficiency for SCIC after discrete cosine transform (DCT). At the encoder, the quantized vectors or transformation coefficients are losslessly compressed by arithmetic coding. At the decoder, the decoded vectors are utilized to restore input image by image decoder network (IDN). Both of SCIC and DNNC frameworks are built on auto-encoder architecture, whose encoder is the RSN network and whose decoder is the IDN network. The encoder is used to condense input's dimensionality inside the networks. Meanwhile, quantization reduces dimensionality in some dimensional space, no matter whether re-sampled vectors is processed by DCT transform. The decoder of auto-encoder architecture reproduces the input image from these quantized vectors. The difference between SCIC and DNNC mainly comes from whether classical transformation such as DCT transform is explicitly applied to reduce statistical correlation of re-sampled vectors.

Obviously, the main difference between our SCIC and \cite{r25} is that our VCN network bridges the huge gaps of gradient back-propagation between RSN and IDN caused by quantization function's non-differentiability. Another difference lies in that our IRSC method is not restricted to image compression at very low bit-rate. Because our VCN network could well back-propagate gradient from decoder to encoder, our method could conduct full-resolution image re-sampling. The third important difference is that our IRSC method can be applied into DNNC framework. Although our IRSC as well as \cite{r27, r28, r29, r30, r31, r32, r33} can process non-differentiability of quantization function for image compression, our IRSC method's application is not restricted to the application of DNN-based image compression.

The rest of this paper is arranged as follows. Firstly, we review traditional post-processing methods and neural network-based artifact removal techniques, as well as image compression methods with DNN in Section 2. Then, we introduce the proposed method in Section 3, which is followed by experimental results in the Section 4. At last, we give a conclusion in the Section 5.

\section{Related work}
We firstly give a review of traditional artifact removal methods, where loop filtering and post-processing filtering are included. Then, we look back on several state-of-the-art artifact removal approaches based on neural network. At last, we give an overview of image compression methods with DNN.

\subsection{Traditional artifact removal approaches}
Within coding loop, loop filtering can be explicitly embedded to improve coding efficiency and reduce artifacts caused by the coarse quantization. For example, adaptive de-blocking filtering \cite{r6} is designed as a loop filter and integrated into H.264/MPEG-4 AVC video coding standard, which does not require an extra frame buffer at the decoder. The priority of de-blocking filtering inside coding loop lies in guaranteeing that an established level of image quality is coded and conveyed in the transmission channel. However, this kind of filtering always has comparatively high computational complexity. Meanwhile, loop filtering should be done at the decoder so as to be synchronized with the encoder, which prevents from adaptively decoding via turning on/off loop filtering, when making a balance between visual quality and computational cost.

In order to avoid these drawbacks and make filtering compatible to standard codec, the alternative flexible manner is to do post-processing. For instance, a wavelet-based algorithm uses three-scale over-complete wavelet to de-block via a theoretical analysis of blocking artifacts \cite{r7}. Later, through image's total variation analysis, adaptive bilateral filters is used as a de-blocking method to process two different kinds of regions \cite{r8}. In contrast, by defining a new metric to evaluate blocking artifacts, quantization noise on blocks is removed by non-local means filtering \cite{r9}. The above methods target at de-blocking. However, coarse quantization on block-based DCT domain usually causes visually unpleasant blocking artifacts as well as ringing artifacts. Thus, both de-blocking and de-artifacts should be carefully considered for better visual quality. In \cite{r10}, both hard-thresholding and empirical Wiener filtering are carried on shape adaptive DCT for de-noising and de-blocking.

Unlike the above mentioned methods \cite{r6, r7, r8, r9, r10}, many works have incorporated some priors or expert knowledge into their models. In \cite{r13}, compression artifacts are reduced by integrating quantization noise model with block similarity prior model. In \cite{r14}, maximum a posteriori criterion is used for compressed image's post-processing by treating post-processing as an inverse problem. In \cite{r15}, an artifact reducing approach is developed by dictionary learning and total variation regularization. In \cite{r17}, image de-blocking is formulated as an optimization problem with constrained non-convex low-rank model. In \cite{r18}, both JPEG prior knowledge and sparse coding expertise are combined for JPEG-compressed images. In \cite{r36}, sparse coding process is carried out jointly in the DCT and pixel domains for compressed image's restoration. Image de-noising is a more general technique, which is not designed for specific task. It can be applied for removing additive Gaussian noise, environment noise, and compression artifact, etc. For example, an advanced image de-noising strategy is used to achieve collaborative filtering based on a sparse representation on transform domain \cite{r11}. In \cite{r38}, self-learning based image decomposition is applied for single image de-noising with an over-complete dictionary, which can be used to alleviate coding artifacts. Although the above methods have good performances on artifacts removal, they always have a fairly high computational complexity via iterative optimization algorithms, which are time-consuming.
\begin{figure*}[ht]
\centering
\includegraphics[width=7in]{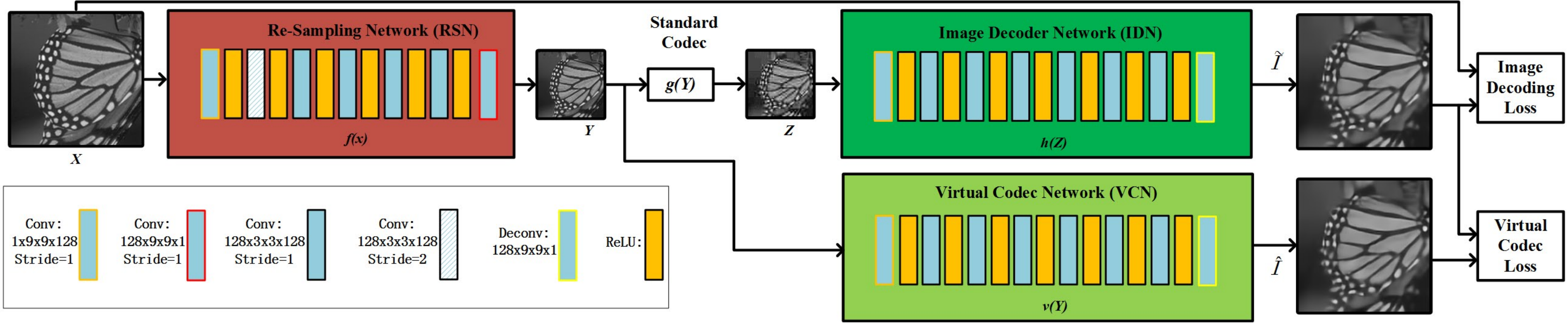}
\caption{The diagram of standard-compliant coding framework with low-resolution re-sampling}
\label{Fig2}
\end{figure*}
\subsection{CNN-based post-processing for standard compression}
Due to neural network's strong capacity, it has been successfully applied for some low-level tasks: such as image super-resolution, image smoothing and edge detection \cite{r40}. With this trend, many works such as \cite{r19, r20, r21, r22, r23, r24, r25} put their research on the issue of CNN-based post-processing to improve user's visual experience. In \cite{r19}, artifacts reduction convolutional neural network is presented to effectively deal with various compression artifacts. To get better results, a 12-layer deep convolutional neural network with hierarchical skip connections is trained with a multi-scale loss function \cite{r21}. Meanwhile, a deeper CNN model is used for image de-blocking to obtain more improvements \cite{r22}. However, these methods are trained by minimizing mean square error, so the reconstructed image usually loses detail at the high frequencies and may be blurry around visually sensitive discontinuities. In order to generate more details, a conditional generative adversarial framework is trained to remove compression artifacts and make generated image very realistic as much as possible \cite{r24}.

Although the above methods have greatly alleviated the problem of ringing artifacts and blocking artifacts, their improvements are usually limited. This raises a new question: whether it's possible to compactly represent image so that codec can more efficiently compress these images. The pioneering work \cite{r25} resolves this problem by directly training two collaborative neural networks: compact convolutional neural network and reconstruction convolutional neural network. This method performs well at very low bit-rate, but it doesn't consider how to resolve this problem at high bit-rate, which strictly restricts their method's wide applications.

For video coding's post-processing, there are several latest works about this issue, such as \cite{r20,r23}. For example, deep CNN-based decoder is presented to reduce coding artifacts and enhance the details of HEVC-compressed videos at the same time \cite{r20}. In \cite{r23}, a convolutional neural network with scalable structure is used to reduce distortion of I and B/P frames in HEVC for quality enhancement. Despite that these approaches greatly reduce coding artifacts by post-processing, these methods \cite{r19, r21, r22, r24, r20, r23} are limited, since their inputs directly use natural images/videos without compactly representing them, when comparing with \cite{r25}.

\subsection{Deep neural networks based image compression}
To achieve variable-rate image compression, a general framework is presented based on convolutional and de-convolutional LSTM recurrent networks \cite{r27}. This method can address 32x32 thumbnails compression, but it may not be suitable for lossy image compression with full-resolution. To resolve this problem, the authors carefully design a full-resolution lossy image compression method, which is composed of a recurrent neural network-based encoder and decoder, a binarizer, and a neural network for entropy coding \cite{r28}. At the same period, nonlinear transform coding optimization framework is introduced to jointly optimize their entire model in terms of the trade-off between coding rate and reconstruction distortion \cite{r29, r30}. Later, a compressive auto-encoder architecture is efficiently trained for high-resolution images using convolutional sampling layer and sub-pixel convolution \cite{r31}. After that, using the same neural network architecture, soft assignments with soft-max function is leveraged to softly relax quantization so as to optimize the rate-distortion loss \cite{r32}. In the meanwhile, bit-rate is allocated for image compression by learning a content weighted importance map and this map is used as a continuous estimation of entropy so as to control image compression's bit-rate. Although these methods have greatly improve coding efficiency, their compressed image always doesn't have pleasing details, especially at very low bit-rate.

Due to generative model's huge progress, image generation becomes better and better. Particularly, generative adversarial networks (GAN) has been widely researched and achieved more stable results than previous methods for image generation, style transfer, and super-resolution, etc \cite{r40}. Following this trend, adversarial loss is introduced into adaptive image compression approach so as to achieve visually realistic reconstructions \cite{r34}. Most recently, semantic label map is leveraged as supplementary information to help GAN's generator to produce more realistic images, especially at extremely low bit-rate \cite{r35}. Although image compression based on DNN has made great progress in some respect, there is still a lot of development space for image/video compression. More importantly, a general image compression method with DNN is required for both standard-compliant image compression and DNN-based image compression.

\section{Methodology}

Given an input image $\bm{X} \in \mathbb{N}^{M \times N}$, we use RSN network to get re-sampled vectors $\bm{Y}$ in the low-dimension space. For the sake of simplicity, RSN network is expressed with a non-linear function $f(\bm{X},\alpha)$, whose parameter set is denoted as $\alpha$. After re-sampling, these vectors are quantized, which is described as a mapping function $\bm{Z}=q(\bm{Y},\beta)$, where $\beta$ is the parameter of quantization. This function will be detailed later. The quantized vectors $\bm{Z}$ are losslessly encoded by arithmetic coding to facilitate channel transmission. Because the vectors $\bm{Z}$ lose some information caused by quantization, there are coding distortions between input image $\bm{X}$ and decoded image $\bm{\tilde{I}}$. At the receiver, IDN network parameterized by $\gamma$ learns a non-linear function $\bm{\tilde{I}}=h(\bm{Z},\gamma)$ to restore the input image from $\bm{Z}$.

Since quantization function is non-differentiable, we can find that this function can't directly optimized by gradient descent method. Several approaches \cite{r27, r28, r29, r30, r31, r32, r33, r34, r35} give their solutions for this problem. Different from these approaches, we learn a approximation function from the re-sampled vectors $\bm{Y}$ to the decoded image $\bm{\tilde{I}}$ with the VCN network, and thus we can use its derivation to approximate the quantization function's derivation during back-propagation. As a consequence, we can optimize our RSN network and IDN network in an end-to-end fashion with the learned VCN network. In order to verify the proposed method's generalization, we use this method for SCIC framework and DNNC framework, which will be detailed next. Note that we employ function $g(\cdot)$ in Fig.\ref{Fig2} rather than directly using quantization function $q(\cdot)$ in Fig.\ref{Fig1}. In the SCIC framework, $g(\cdot)$ represents the mapping function from the re-sampled vector to the decoded lossy re-sampled vector through several steps: transformation such as DCT transform, quantization, arithmetic coding, de-quantization and inverse transformation.

\subsection{Standard-compliant image compression framework}
To make our framework suitable for different scenarios, we use mix-resolution image compression in this framework, so that our method has high coding efficiency ranging from low bit-rate to high bit-rate. Specifically, full-resolution re-sampling for RSN network is designed for image compression at high bit-rate. When compressing image below certain low bit-rate, that is, each pixel's quality is very low, image can't be well restored from full-resolution re-sampled vectors due to each pixel having little bits to be assigned. Furthermore, there is almost no more bit assigned for image details, so only image structures are mainly kept after decoding. Therefore, down-sampling layer for RSN network is leveraged to greatly reduce image information, so that each pixel of low-resolution re-sampled image can be assigned with more bits, as compared to full-resolution re-sampled vectors. As a result, in relative to full-resolution, we can get high-quality but low-resolution images at the decoder, which are leveraged to restore high-quality yet full-resolution image by IDN network. The details about how to choose low-resolution re-sampling or full-resolution re-sampling will be presented in the experimental section.

\subsubsection{Objective function}
Our objective compressive function for SCIC framework can be written as follows:
\begin{align}
&\mathop{\arg\min}_{\alpha, \gamma, \theta} L_{IDN}(\bm{\tilde{I}, \bm{X}})+ L_{VCN}(\bm{\hat{I}},\bm{\tilde{I}})+L_{DSSIM}(s(\bm{Y}),\bm{X}),\notag\\
&\bm{Y}=f(\bm{X},\alpha),\bm{\tilde{I}}=h(\bm{Z},\gamma), \bm{Z}=g(\bm{Y},\beta), \bm{\hat{I}}=v(\bm{Y},\theta),
\end{align}
where the first term is image decoding loss for IDN network, the second term is virtual codec loss for VCN network. Meanwhile, the last term is structural dis-similarity (DSSIM) loss, which is explicitly used to regularize RSN network. Here, RSN, IDN, and VCN are parameterized by $\alpha$, $\gamma$, and $\theta$ respectively, while $s(\cdot)$ is the linear up-sampling operator so that $s(\bm{Y})$ could keep the same image size with $\bm{X}$ for low-resolution re-sampling. But, $s(\bm{Y})=\bm{Y}$, when image compression takes full-resolution re-sampling.

In order to decode image $\bm{\tilde{I}}$ with IDN network from $\bm{Z}$, as shown in Fig.\ref{Fig2}, we use data loss and gradient difference loss to regularize IDN network. Meanwhile, our VCN network is trained with data loss as well as gradient different loss between $\bm{\tilde{I}}$ and $\bm{\hat{I}}$. It has been reported that the L1 norm has better performance to supervise convolutional neural network's training than L2 norm \cite{r16} and \cite{r40}. For example, future image prediction from video sequences is learned via loss function with L1 norm \cite{r16}. Moreover, both gradient difference loss and data loss with L1 norm are used to supervise tasks of simultaneous color-depth image super-resolution or concurrent edge detection and image smoothing with conditional GAN \cite{r40}. Accordingly, we use the L1 norm for our data loss and gradient difference loss. Data loss can be defined as:
\begin{equation}
\begin{split}
L_{data}(\bm{A},\bm{B})= \frac{1}{M \cdot N}\sum_{i}(||\bm{A}_i-\bm{B}_i||_1).
\end{split}
 \label{eqn::data loss}
\end{equation}
Gradient difference loss can be written as:
\begin{equation}
\begin{split}
L_{gradient}(\bm{A},\bm{B})= \frac{1}{M \cdot N} \sum_{i} (\sum_{k\in{\Omega}}||\nabla_k \bm{A}_i-\nabla_k \bm{B}_i||_1),
\end{split}
\label{eqn::GRADIENT}
\end{equation}
where $\nabla_k$ is the $k$-th gradient between each pixel and $k$-th pixels among its 8-neighbourhood $\Omega$.

Usually, it's hoped that decoded re-sampled vectors are able to be watched by the receivers, even though without IDN network's processing, so the re-sampled vector's structural information should be similar to the input image $\bm{X}$. As a consequence, $L_{DSSIM}(s(\bm{Y}),\bm{X})$ is used to further supervise the learning of RSN network, in addition to loss from the IDN network. Based on \cite{r39}, DSSIM loss between $s(\bm{A})$ and $\bm{B}$ can be defined as follows:
\begin{equation}
\begin{split}
L_{DSSIM}(s(\bm{A}),\bm{B})=1-\frac{1}{M \cdot N} \sum_{i} L_{SSIM}(s(\bm{A})_i,\bm{B}_i)
\end{split}
\label{eqn::SSIMLOSS}
\end{equation}
\begin{align}
&L_{SSIM}(s(\bm{A})_i,\bm{B}_i)=\notag\\
&\frac{(2\mu_{s(\bm{A})_i}\cdot \mu_{\bm{B}_i}+c1)(2\sigma_{s(\bm{A})_i \bm{B}_i}+c2)}{(\mu^2_{s(\bm{A})_i}+\mu^2_{\bm{B}_i}+c1)(\sigma^2_{s(\bm{A})_i}+\sigma^2_{\bm{B}_i}+c2)}
\end{align}
where $c1$ and $c2$ are two constants. We set them respectively to be $0.0001$ and $0.0009$. $\mu_{\bm{B}_i}$ and $\sigma^2_{\bm{B}_i}$ respectively are mean value and variance of the neighborhood window centered by pixel $i$ in the image $\bm{B}$. Similarly, $\mu_{s(\bm{A})_i}$ as well as $\sigma^2_{s(\bm{A})_i}$ can be denoted in this way. Meanwhile, $\sigma_{s(\bm{A})_i \bm{B}_i}$ refers to the covariance between neighborhood windows centered by pixel $i$ in in the image $s(\bm{A})$ and in the image $\bm{B}$. As we all know, the function of structural similarity index (SSIM) is differentiable, so $L_{DSSIM}(s(\bm{A}),\bm{B})$ can be optimized with gradient descent method. Besides, DSSIM loss between $\bm{\hat{I}}$ and $\bm{\tilde{I}}$ is explicitly used to regularize the VCN network, when down-sampling layer is employed in the RSN network, since the mapping from the re-sampled vectors to the compressed lossy image should be well learned for efficient back propagation at very low bit-rate.

\subsubsection{Network}
In the RSN network, seven convolutional layers are used to re-sample $\bm{X}$ to get $\bm{Y}$, as shown in Fig.\ref{Fig2}. Within this network, the weight's spatial size of these convolutional layers is 9x9 in the first layer and last layer, which makes convolutional neural network's receptive field large. Other five convolutional layers in the RSN network use 3x3 convolution kernel to further enlarge the size of receptive field. These convolutional layers are used to increase the non-linearity of the network, when ReLU is followed to activate the output features of these convolutional layers. The output feature map number of 1-6 convolutional layers is 128, but the last layer only has one output feature map so as to keep consistent with the input image $\bm{X}$. Each convolutional layer is operated with a stride of 1, except that the second layer uses stride step of 2 to down-sample feature maps, so that the convolution operation is carried out in the low-dimension space to reduce computational complexity from the third convolutional layer to the 7-th convolutional layer. It's worthy to noticing that the second layer uses stride step of 1 so that $\bm{Y}$ is re-sampled with full-resolution, when the coding bit-rate is beyond certain values. All the convolutional layers are followed by an activation layer with ReLU function, except the last convolutional layer.

In the IDN network, we leverage seven convolutional layers to extract features and each layer is activated by ReLU function. The size of convolutional layer is 9x9 in the first layer and the left six layers use 3x3, while the output channel of feature map equals to 128 in these convolutional layer. After these layers, one de-convolution layer with size of 9x9 and stride to be 2 is used to up-scale feature map from low-resolution to high-resolution for low-resolution re-sampling compression. Thus, the size of output image is matched with the ground truth image. However, if $\bm{Y}$ is full-resolution, the last de-convolution layer is replaced by convolutional layer with size of 9x9 and stride to be 1.

In our method, VCN network is designed to have the same network structure as IDN network, because they are the same class of low-level image processing problems. The role of the VCN network is to make the re-sampled vectors $\bm{Y}$ degrade to a decoded lossy but full-resolution image $\bm{\hat{I}}$. Different from VCN network, IDN network works to restore input image from quantized re-sampled vectors $\bm{Z}$ so that the user could receive a high-quality image $\bm{\tilde{I}}$ at decoder.

\subsection{Deep neural networks based compression framework}
Here, we choose the auto-encoder architecture in \cite{r31} for DNNC framework, but the sub-pixel convolutional layer for color image is replaced by de-convolutional layer for gray image compression. The encoder network within our framework is called RSN network, while decoder network is named IDN network. From \cite{r31}, it can be easily found that the major components of auto-encoder architecture are convolutional layer with stride to be 2 and sub-pixel convolution as well as ResNet blocks.

Previous approaches such as \cite{r27,r28,r29,r30,r31,r32,r33,r34,r35} use the specific approximation functions to make quantization differentiable so that make their framework trained in an end-to-end way. For example, One direct way is to replace quantization function with differentiable quantization such as stochastic rounding function \cite{r27,r28}, soft-to-hard quantization \cite{r32}, or replace quantization step with additive uniform noise \cite{r29, r30}. The other alternative way is only to use approximation function's gradient during back-propagation \cite{r31}, but the forward pass still uses classic quantization in order to not change gradients of the decoder network.

We provide a novel way to resolve this problem, which is to learn virtual codec (VCN network), and thus the gradient of quantization function from the IDN network to RSN network can be approximated by the VCN network's gradient during back-propagation. The objective compressive loss function can be defined as:
\begin{equation}
\mathop{\arg\min}_{\alpha, \gamma, \theta} L_{IDN}(\bm{\tilde{I}, \bm{X}})+ L_{VCN}(\bm{\hat{I}},\bm{\tilde{I}}),
\end{equation}
\begin{equation}
L_{IDN}(\bm{\tilde{I}, \bm{X}})=L_{data}(\bm{\tilde{I}, \bm{X}}), L_{VCN}(\bm{\hat{I}},\bm{\tilde{I}})=L_{data}(\bm{\hat{I}},\bm{\tilde{I}})
\end{equation}
\begin{equation}
\bm{Y}=f(\bm{X},\alpha),\bm{\tilde{I}}=h(\bm{Z},\gamma), \bm{Z}=q(\bm{Y},\beta), \bm{\hat{I}}=v(\bm{Y},\theta), 
\end{equation}
in which the symbol marking is similar to Eq.(1). Here, $L_{IDN}(\bm{\tilde{I}, \bm{X}})$ is the image decoding loss for IDN network and $L_{VCN}(\bm{\hat{I}},\bm{\tilde{I}})$ is the virtual codec loss for VCN network.

\begin{figure}[t]
\centering
\includegraphics[width=3.5in]{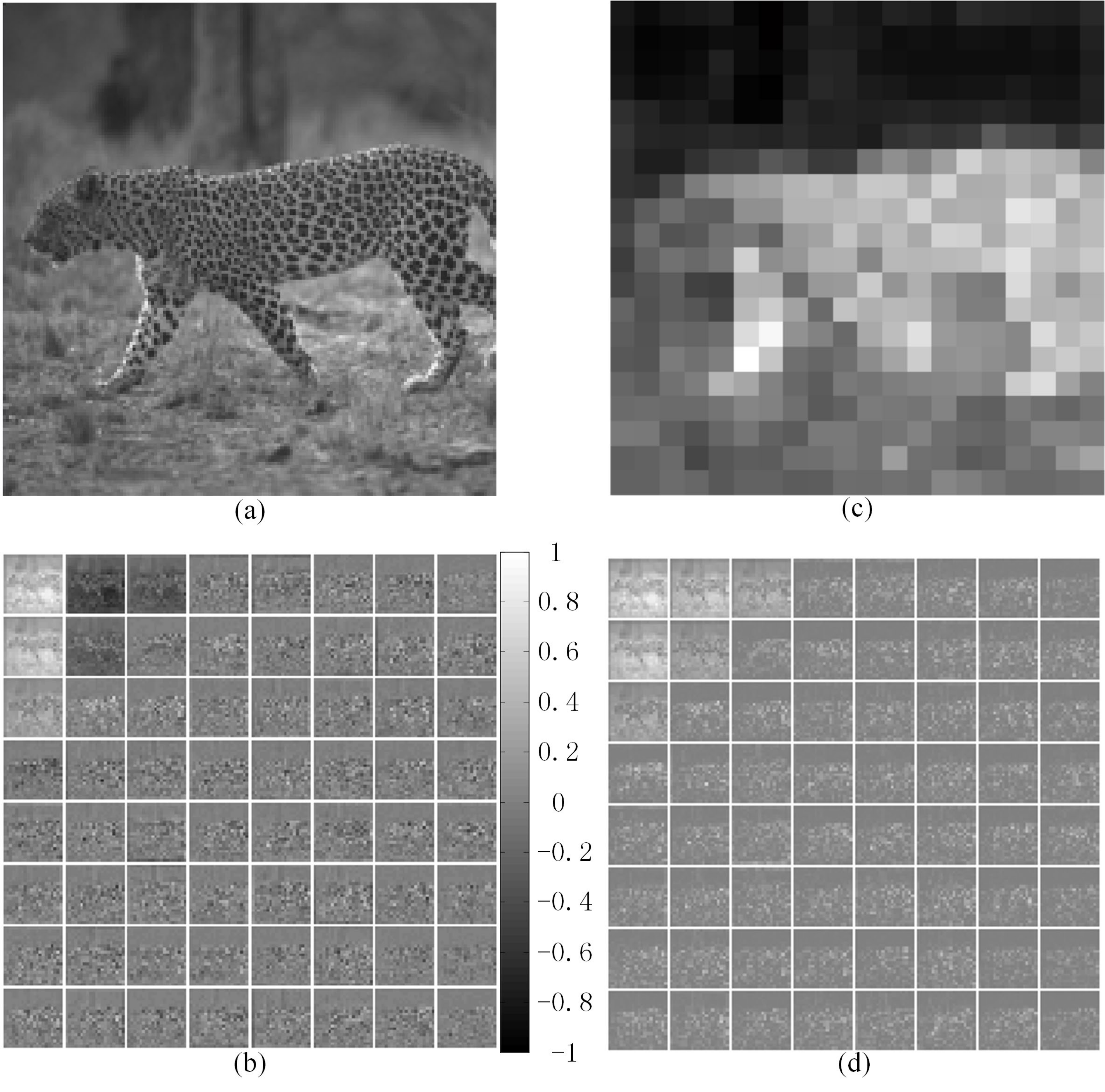}
\caption{The re-sampled vectors for DNNC framework (a) input image, (b) re-sampled vectors listed in Zig-zag scanning order, (c) pixel-wise sum of all the re-sampled vectors with absolute values, (d) re-sampled vectors with absolute values}
\label{Fig3}
\end{figure}

Different from Eq.(1), there is no DSSIM loss between $s(\bm{Y})$ and $\bm{X}$, because the re-sampled vectors are hoped to work like wavelet transform, which decomposes input image $\bm{X}$ to low-frequency components and high-frequency components. Thus, we don't impose a SSIM loss restriction on re-sampled vectors within the DNNC framework. As shown in Fig.\ref{Fig3}, the re-sampled vectors are listed in the zig-zag scanning order, from which we can see that the RSN network decompose $\bm{X}$ into multiple components. Each component contains particular information of image $\bm{X}$. We can well restore the input image from these vectors, when these vectors are losslessly transmitted over the channel. In order to further compress these vectors, quantization can be operated on these vectors and then the quantized vectors are encoded by arithmetic coding. Without learning parameters of quantization, we first normalize the re-sampled vectors between 0 and 1. And then we re-scale and round them to be integers among $[0, \beta]$. They can be written as:
\begin{align}%
&\ddot{Y}(\bm{Y},\beta)=round(\beta*(\bm{Y}-\bm{Y}_{min}/(\bm{Y}_{max}-\bm{Y}_{min}))),\notag\\
&\bm{Y}_{min} < 0, \bm{Y}_{max} > 0,
\end{align}
where $\bm{Y}_{min}$ and $\bm{Y}_{max}$ is the minimum value and the maximum value of $\bm{Y}$ among training data's re-sampled vectors using the pre-trained network. Accordingly, $\bm{Z}=q(\bm{Y},\beta)$ can be written as:
\begin{equation}
\begin{split}
\bm{Z}=q(\bm{Y},\beta)=\ddot{Y}(\bm{Y},\beta)/\beta*(\bm{Y}_{max}-\bm{Y}_{min})+\bm{Y}_{min}.
\end{split}
\end{equation}

When we set feature map's number to be constant value like \cite{r31}, the re-sampled vectors always tend to have some redundancy, which leads to high bit-rate coding. Thus, we change the feature map's numbers $\mathcal{N}$ to control image compression's bit-rate, e.g., $\mathcal{N}=1, 2, 4, 8, 12,  16, 20$ for compact image re-sampling. Meanwhile, we set quantization parameter $\beta$ to constant value $64$ in our DNNC framework, from which we can see that our DNNC framework doesn't require to learn the quantization parameter.

\subsection{Learning algorithm for both of our frameworks}
\begin{algorithm}[t]
\caption{Learning Algorithm for Image Compression with Virtual Codec Supervised Re-Sampling Network}
\scriptsize
\begin{algorithmic}[1]
\renewcommand{\algorithmicrequire}{\textbf{Input:}}
\renewcommand{\algorithmicensure}{\textbf{Output:}}
\Require Ground truth image: $\bm{X}$; the number of iteration: $K$; the total number of images for training: $n$; the batch size during training: $m$;
\Ensure  The parameter sets of RSN network and IDN network: $\alpha$, $\gamma$;
\State After auto-encoder networks are firstly pre-trained, the RSN network is initialized with the encoder of this auto-encoder. Meanwhile, the decoder is used to initialize the IDN network and VCN network. At the beginning, the re-sampled vectors are got with this initialized IDN network
\State The initialization of parameter sets: $\alpha$, $\beta$, $\gamma$, $\theta$;
\For{$k=1$ to $K$}
    \State The re-sampled vectors are quantized with parameter $\beta$
    \For{$epoch=1$ to $p$}
        \For{$i=1$ to floor$(n/m)$}
            \State Update the parameter set of $\gamma$ by optimizing the IDN network with $i$-th
            \State batch images using gradient descent method
        \EndFor
    \EndFor
    \For{$epoch=1$ to $p$}
        \For{$j=1$ to floor$(n/m)$}
            \State Update the parameter set of $\theta$ by optimizing the VCN network with $j$-th
            \State batch images
        \EndFor
    \EndFor
    \For{$epoch=1$ to $q$}
        \For{$l=1$ to floor$(n/m)$}
            \State Update the parameter set of $\alpha$ with fixing $\theta$ by optimizing RSN network
            \State with $l$-th batch images
        \EndFor
    \EndFor
\EndFor
\State Update the parameter set of $\gamma$ by optimizing the IDN network
\State \textbf{return} $\alpha$, $\gamma$;
\end{algorithmic}
\end{algorithm}

Due to the difficulty of directly training the whole framework once, we decompose the learning of three convolutional neural networks as three sub-problems learning. First, we initialize the parameter sets, $\alpha$, $\gamma$, and $\theta$ of RSN network, IDN network, and VCN network. Because both of our frameworks are built on the auto-encoder, we can initialize these three networks by pre-training auto-encoder networks, which contains RSN network and IDN network without quantization. In fact, our two frameworks become classical auto-encoder networks, when there is no quantization. After the initialization neural networks, we use RSN network to get an initial re-sampled vector $\bm{Y}$ from the input image $\bm{X}$, which is then lossly encoded by standard codec or quantized by rounding function as the training data at the beginning. Next, the first sub-problem learning is to train IDN network by updating the parameter set of $\gamma$. The re-sampled vectors $\bm{Y}$ and IDN-decoded image $\bm{\tilde{I}}$ are used for the second sub-problem's learning of VCN to update parameter set of $\theta$. After VCN's learning, we fix the parameter set of $\theta$ in the VCN network to carry on the third sub-problem learning by optimizing the parameter set of $\alpha$ in the RSN network. After RSN network's learning, the next iteration begins to train the IDN network, after the updated re-sampled vectors are compressed by the standard codec for the SCIC framework or only quantized by rounding function for the DNNC framework. The whole training process is summarized in the \textbf{Algorithm-1}. It is worth mentioning that the functionality of VCN network is to bridge great gap between RSN network and IDN network. Thus, once the training of the whole framework is finished, the VCN network is not in use any more, that is to say, only the parameter sets of $\alpha$, $\gamma$ in the networks of RSN and IDN are used during testing.

\begin{figure}[!ht]
\centering
\includegraphics[width=3.5in]{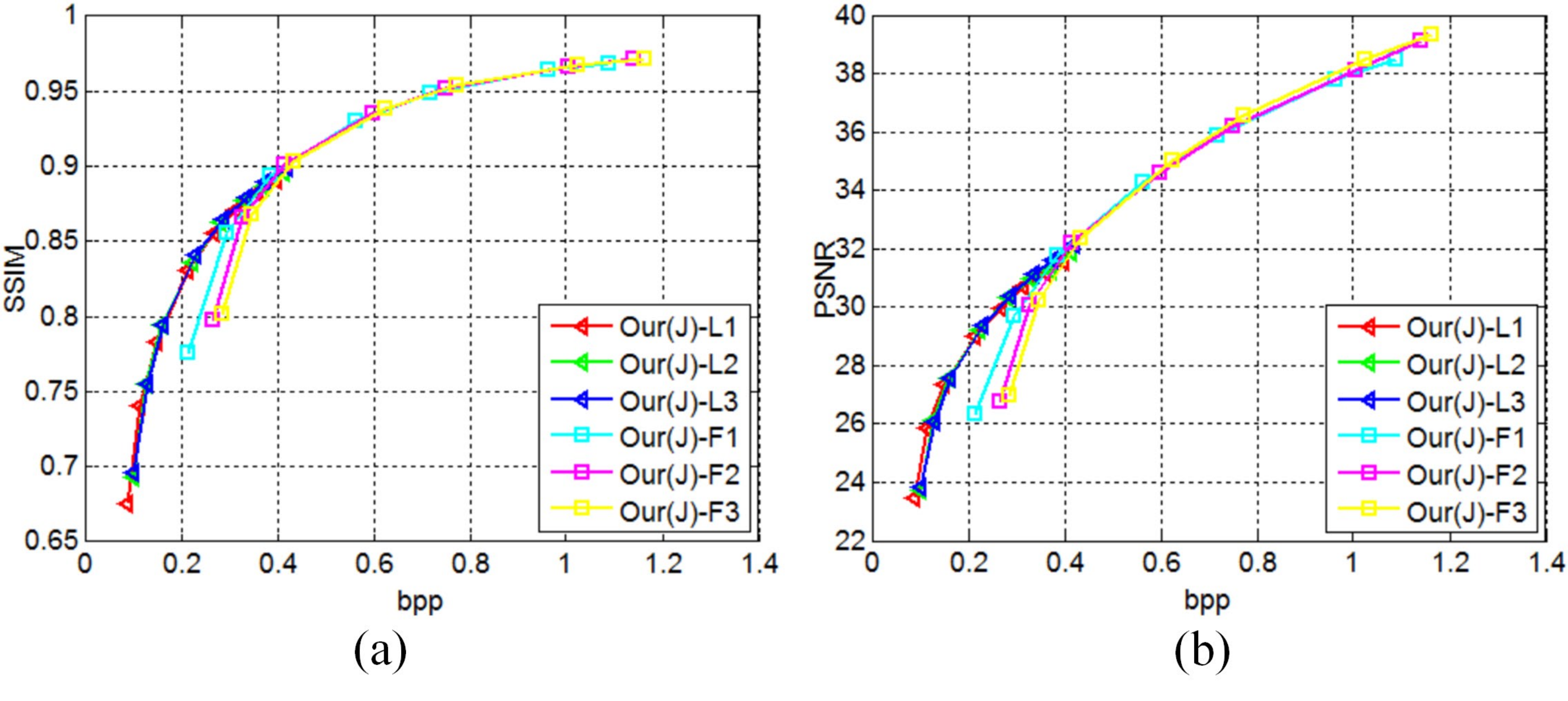}
\caption{The objective quality comparisons of iterative number's effects on the performance in terms of SSIM(a) and PSNR(b), when compressing images on the validation set using our JPEG-compliant image compression with \textbf{Algorithm-1}}
\label{Fig4}
\end{figure}

\section{Experiment and analysis}

To validate the versatility and effectiveness of the proposed method, we apply our image re-sampling compression method for SCIC framework as well as DNNC framework. In our JPEG-compliant image compression framework, namely "Ours(J)", we compare it with JPEG, JPEG2000 and several combinatorial methods, which are standard JPEG compression followed by several artifacts removal, such as \cite{r15}, \cite{r10}, \cite{r17}, \cite{r19}. These combinatorial methods are respectively denoted as "DicTV", "Foi's", "CONCOLOR", "ARCNN". Among these methods, "ARCNN" is the class of CNN-based artifact removal method. Meanwhile, we compare our learning algorithm with one highly related learning algorithm proposed in \cite{r25}. To clearly observe the differences between these two algorithms, we use our RSN network and IDN network trained by the learning algorithm presented in \cite{r25}, whose results are called as "Jiang's". Here, RSN network and IDN network correspond to "ComCNN" and "RecCNN" networks in \cite{r25}. Meanwhile, other learning details of this approach such as training dataset and batch-size, etc., keep consistency with ours, except learning algorithm. In other words, "Jiang's" directly trains RSN network and IDN network in an iterative way, while our method trains VCN network to bridge the gap, which is quantization function's non-differentiability, between RSN network and IDN network. Moreover, we compare our DNNC framework's compression results, which is denoted as Our(D), with JPEG, JPEG2000 and Our(J). Among these comparison, two objective measurements: SSIM, and Peak Signal to Noise Ratio (PSNR) are used to evaluate the efficiency of different image compression methods.

\begin{figure}[!ht]
\centering
\includegraphics[width=3.5in]{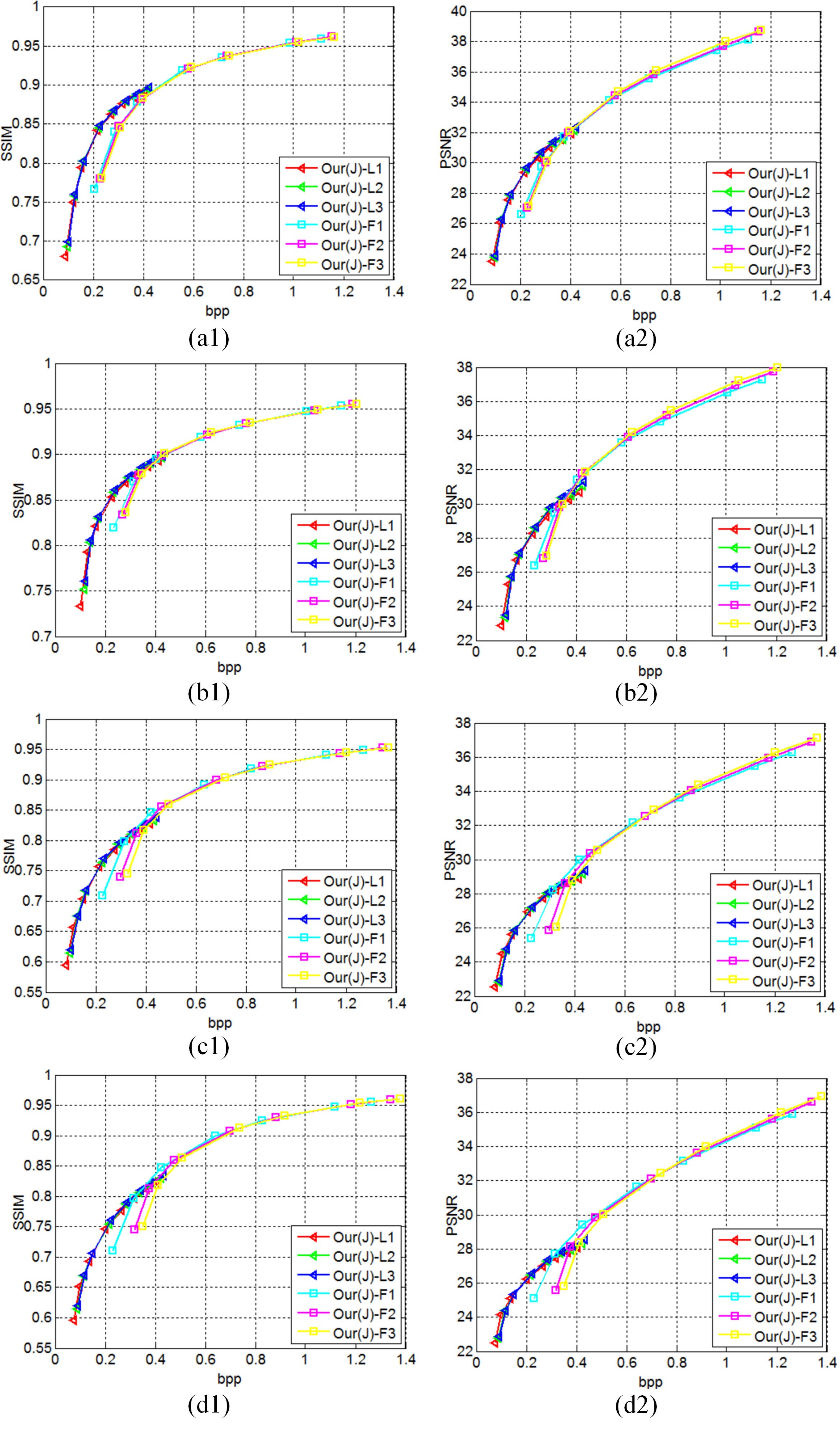}
\caption{The objective quality comparisons of iterative number's effects on the performance in terms of SSIM(a1-d1) and PSNR(a2-d2), when using our JPEG-compliant image compression with \textbf{Algorithm-1}. (a-d) are respectively the average results tested on Set5, Set7, Set14, and LIVE1}
\label{Fig6}
\end{figure}

\subsection{Training details}

To get training dataset, 291 images come from \cite{data1} and \cite{data2}. Among these images, 91\footnote{\url{https://www2.eecs.berkeley.edu/Research/Projects/CS/vision/bsds/}} images are from \cite{data1}, while others\footnote{\url{https://www.ifp.illinois.edu/~jyang29/codes/ScSR.rar}} use BSDS500's training set. Our training dataset consists of 1681 image patches with size 160x160, which are formed by cropping, down-sampling and assembling with small patches, whose size is less than 160x160. During training, each batch of image patches rotates and flips randomly. Moreover, the dataset of General-100 is used as the validation dataset.

To verify the effectiveness of the proposed method, we use several testing dataset: Set5, Set7, Set14, and LIVE1. Among them, the dataset of Set7 is built up with seven testing image by \cite{r25}, while other datasets are widely used for image super-resolution, artifacts removal and image compression. Because some of comparative method mentioned above require image size to be an integer multiple of 8, all the testing images are cropped to be an integer multiple of 8. All the training dataset, validation dataset and testing dataset can be downloaded according to the website \footnote{\url{https://github.com/VirtualCodecNetwork}}.

Our frameworks are implemented in the platform of TensorFlow. Our models are trained using Adam optimization method with beta1=0.9, beta2=0.999. The initial learning rates for three convolutional neural network are set to be 0.0001, while the learning rates decay to be half of last learning rate, once the training step reaches 3/5 and 4/5 of total step.
\begin{figure}[!t]
\centering
\includegraphics[width=3.5in]{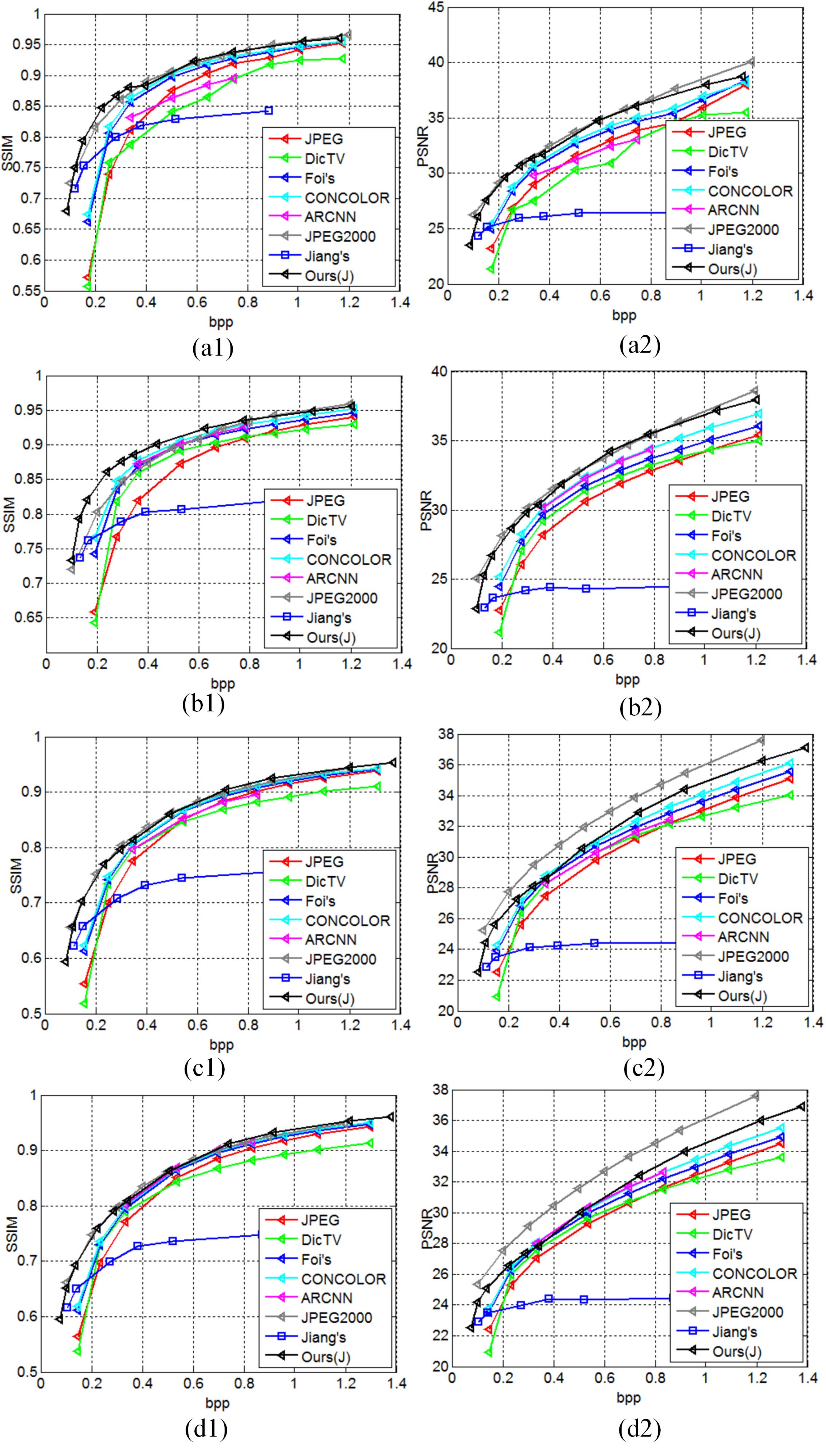}
\caption{The objective quality comparisons of different compression methods in terms of SSIM(a1-d1) and PSNR(a2-d2). (a-d) are respectively the average results tested on Set5, Set7, Set14, and LIVE1}
\label{Fig5}
\end{figure}

\begin{figure}[!ht]
\centering
\includegraphics[width=3.5in]{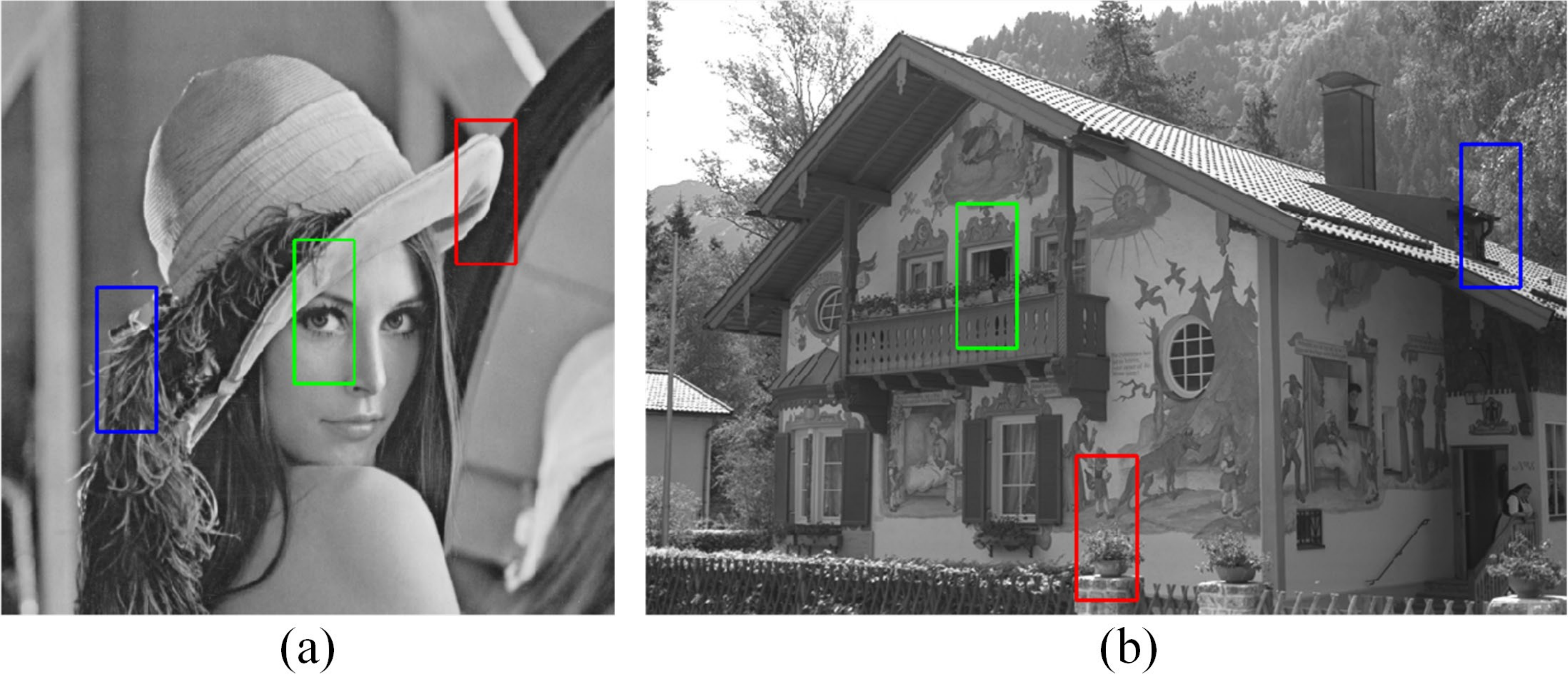}
\caption{Two testing image examples respectively from Set7 and LIVE1 for visual quality comparisons between our JPEG-compliant image compression and several standard codecs}
\label{Fig7}
\end{figure}

\begin{figure*}[!t]
\centering
\includegraphics[width=6.6in]{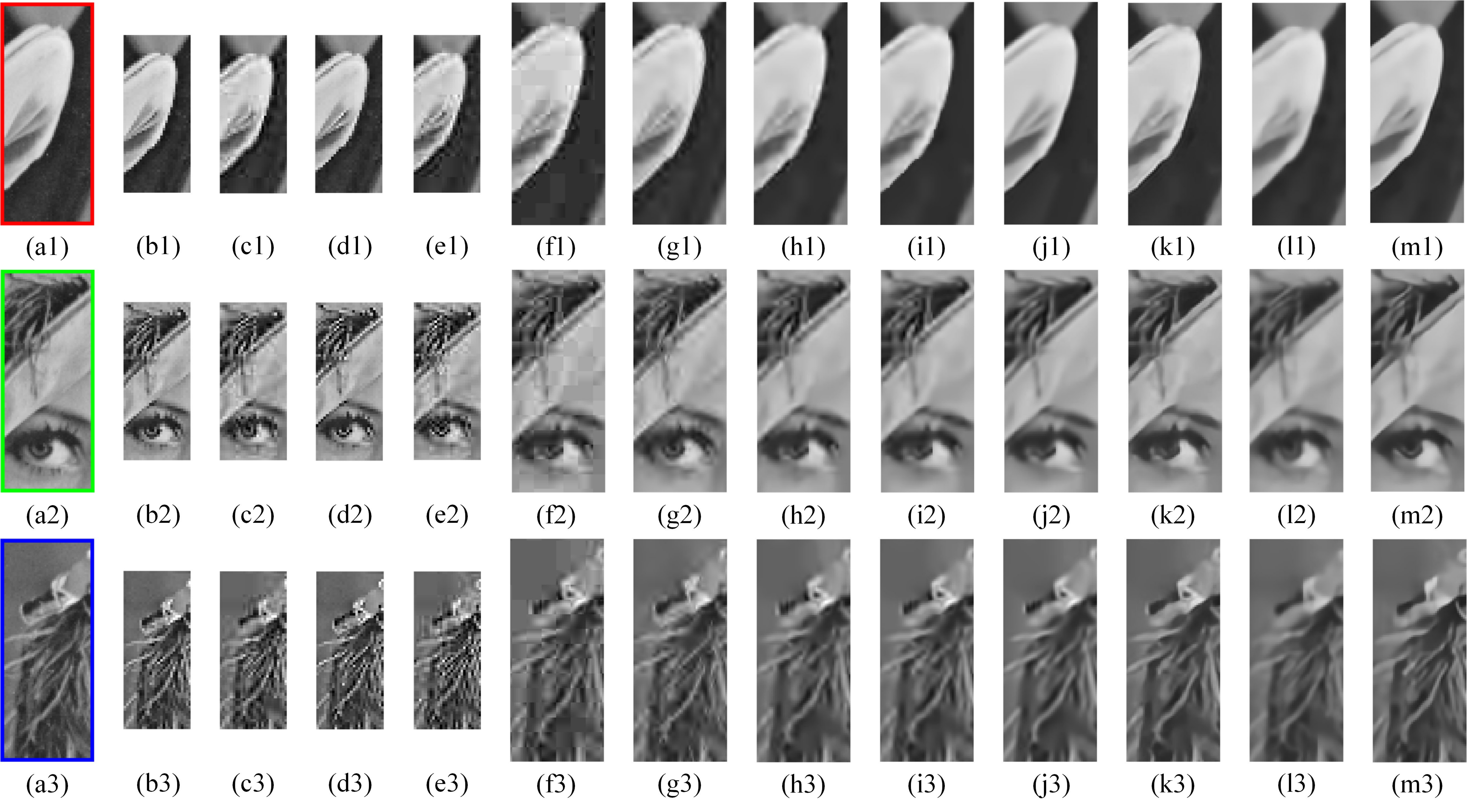}
\caption{The visual quality comparisons between our JPEG-compliant image compression and several standard codecs (a1-a3) enlargements from Fig.\ref{Fig7}(a), (b1-b3) Jiang's compact representation in low-resolution space, (c1-c3) the compressed images of Jiang's compact representation (b1-b3), (d1-d3) our full-resolution sampled images, (e1-e3) the compressed images of our low-resolution sampled images (d1-d3), (f1-f2) JPEG (bpp=0.24), (g1-g3) JPEG2000 (bpp=0.2), (h1-h3) DicTV (bpp=0.24), (i1-i3) Foi's (bpp=0.24), (j1-j3) CONCOLOR (bpp=0.24), (k1-k3) ARCNN (bpp=0.24), (l1-l3) Jiang's (bpp=0.2), (m1-m3) Ours(J) (bpp=0.2)}
\label{Fig8}
\end{figure*}

\begin{figure*}[!t]
\centering
\includegraphics[width=6.6in]{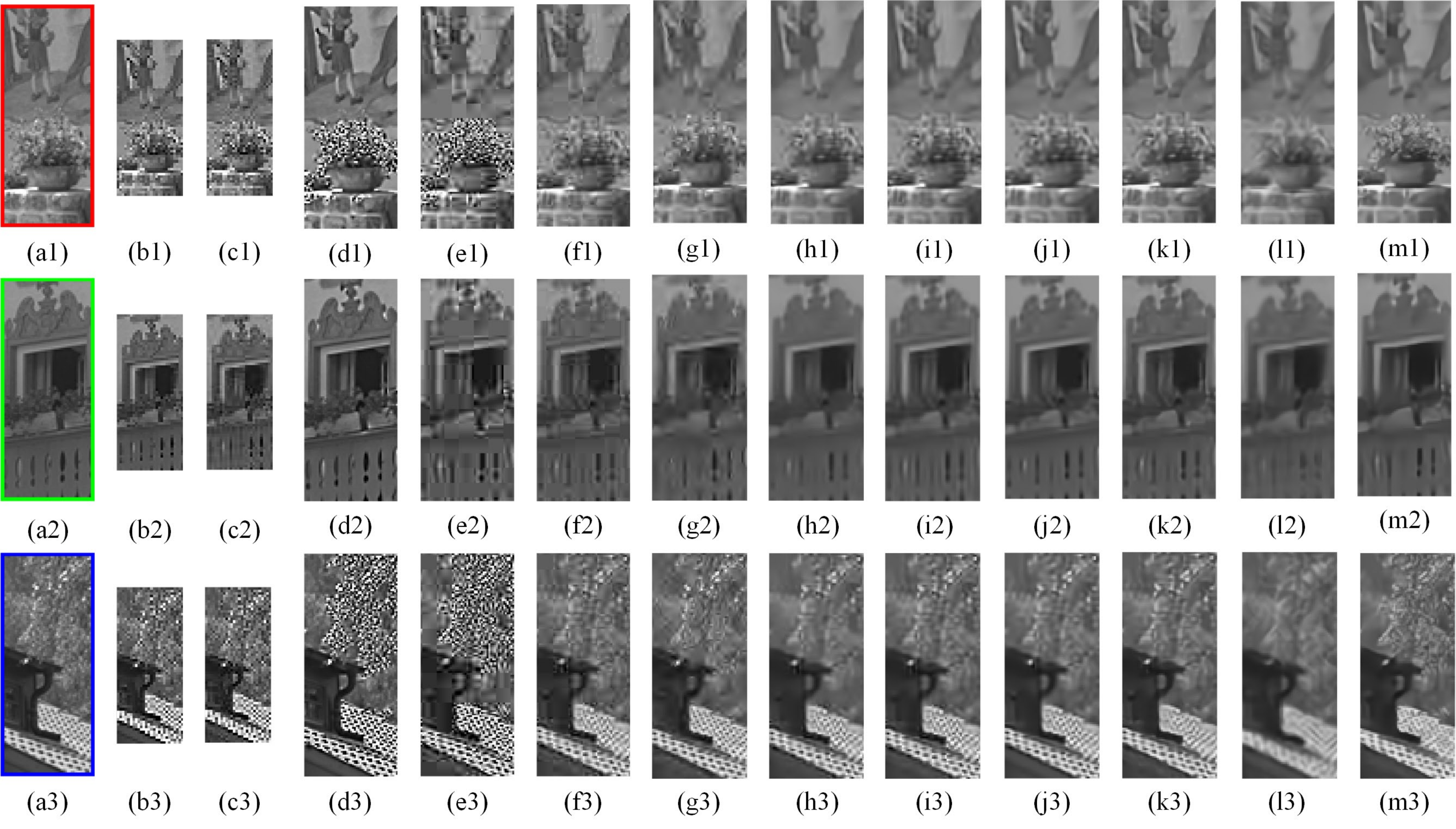}
\caption{The visual quality comparisons between our JPEG-compliant image compression and several standard codecs (a1-a3) enlargements from Fig.\ref{Fig7}(b), (b1-b3) Jiang's compact representation in low-resolution space, (c1-c3) the compressed images of Jiang's compact representation (b1-b3), (d1-d3) our full-resolution sampled images, (e1-e3) the compressed images of our full-resolution sampled images (d1-d3), (f1-f2) JPEG (bpp=0.56), (g1-g3) JPEG2000 (bpp=0.5), (h1-h3) DicTV (bpp=0.56), (i1-i3) Foi's (bpp=0.56), (j1-j3) CONCOLOR (bpp=0.56), (k1-k3) ARCNN (bpp=0.56), (l1-l3) Jiang's (bpp=0.55), (m1-m3) Ours(J) (bpp=0.54)}
\label{Fig9}
\end{figure*}
\subsection{Quantitative and qualitative evaluation between SCIC framework and several state-of-the-art methods}
Our image re-sampling within the SCIC framework not only refers to image full-resolution re-sampling but also image low-resolution re-sampling. Thus, we first need to choose full-resolution re-sampling or image low-resolution re-sampling at some points of bit-per-pixel (bpp). The results of our(J) testing on the validation dataset General-100 with different iterative number as well as different re-sampling ways are shown in Fig.\ref{Fig4}, where Our(J)-L3 and Our(J)-F3 respectively represent our(J) by low-resolution and full-resolution re-sampling using \textbf{Algorithm-1} with $K=3$. Similarly, we denote others in this way, such as Our(J)-L1, Our(J)-L2, Our(J)-F1, and Our(J)-F2.

\begin{figure}[!t]
\centering
\includegraphics[width=3in]{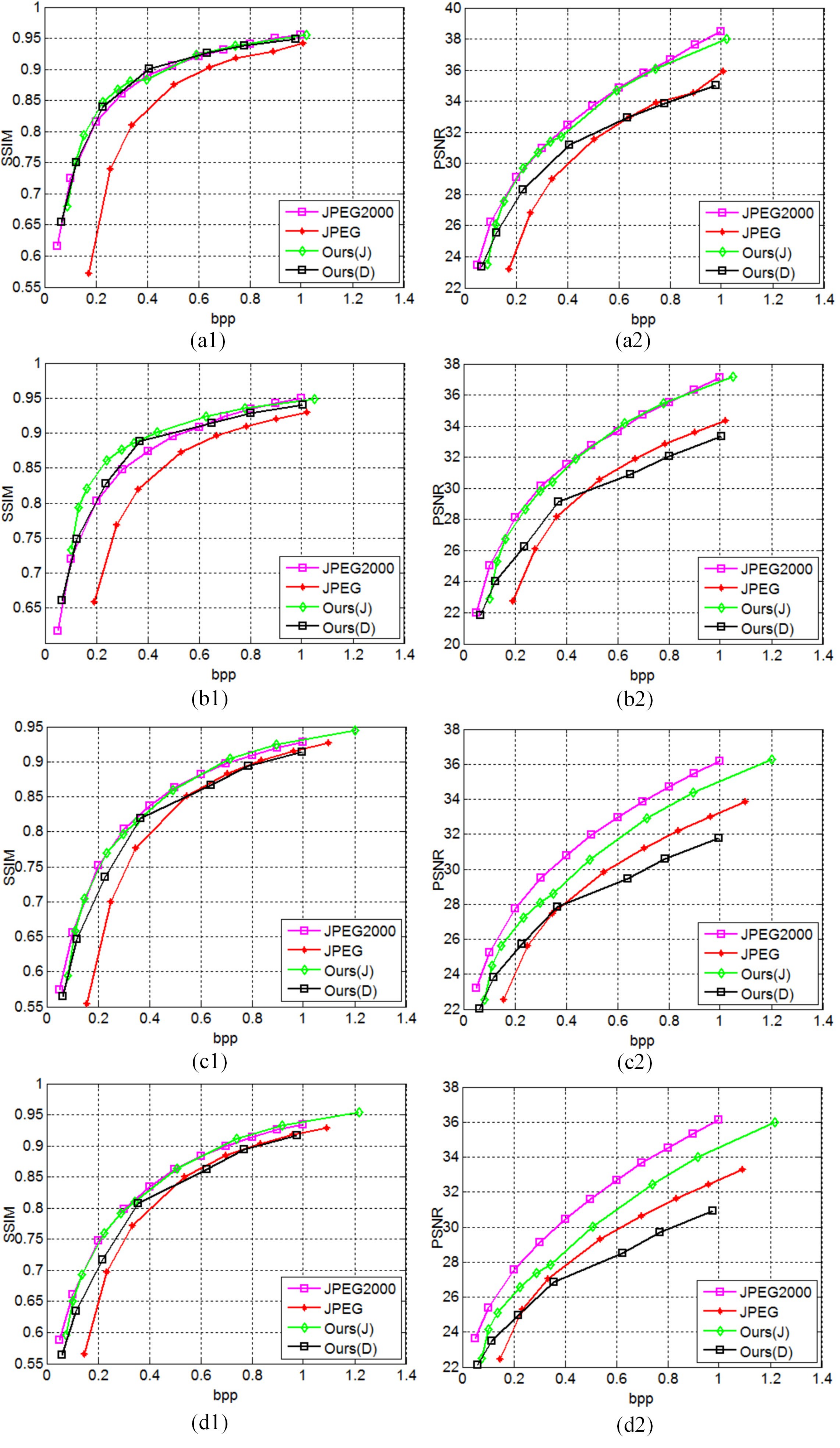}
\caption{The objective quality comparisons for Our(D) and Our(J), JPEG2000 as well as JPEG in terms of SSIM(a1-d1) and PSNR(a2-d2) (a-d) are respectively the average results tested on Set5, Set7, Set14, and LIVE1}
\label{Fig10}
\end{figure}

\begin{figure*}[!t]
\centering
\includegraphics[width=6.5in]{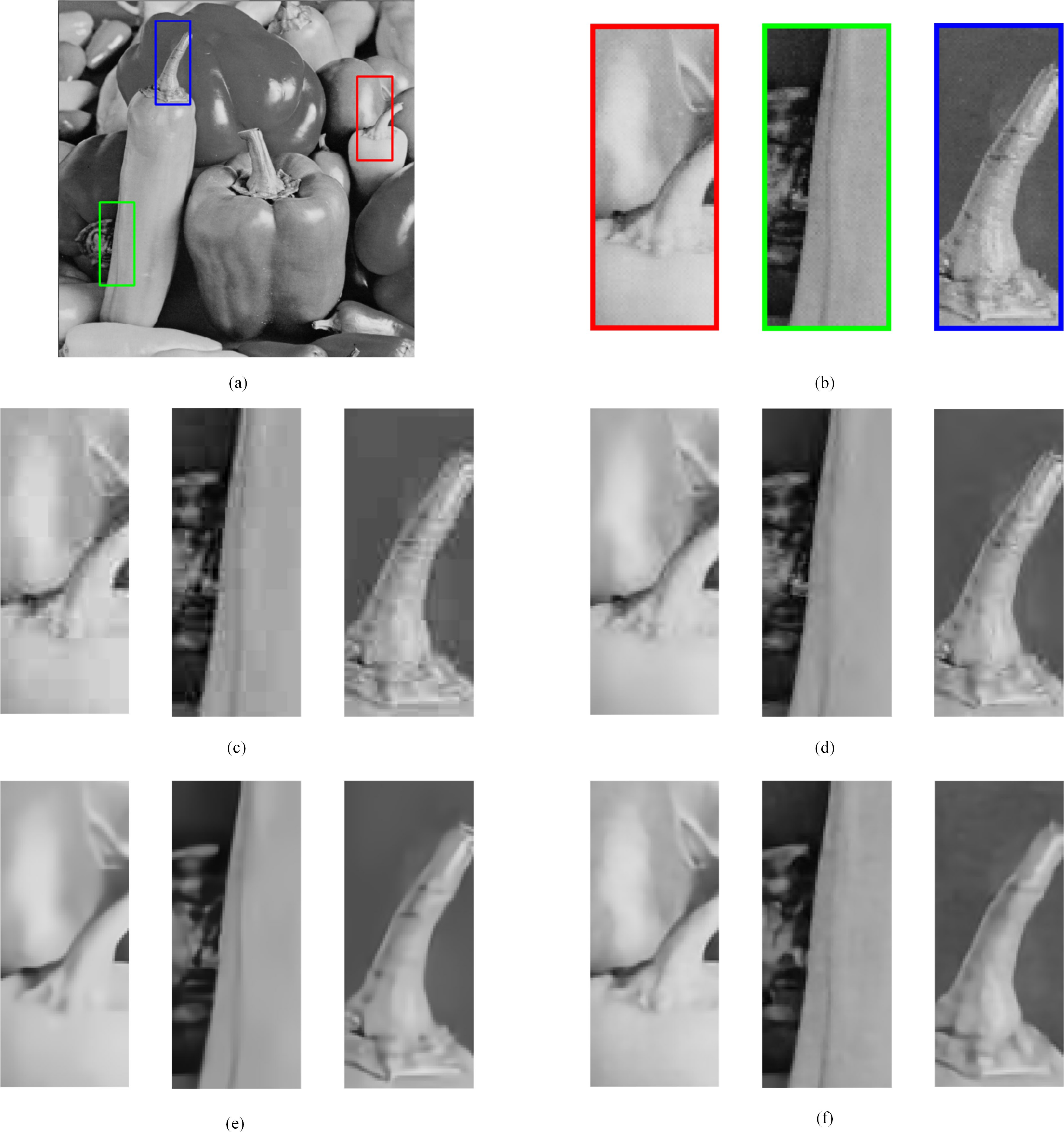}
\caption{The visual quality comparisons for Our(D) and Our(J), JPEG2000 as well as JPEG. (a) Input image, (b) enlargements from (a), (c) JPEG (bpp=0.36), (d) JPEG2000 (bpp=0.33), (e) Ours(J) (bpp=0.28), (f) Ours(D) (bpp=0.33)}
\label{Fig11}
\end{figure*}
\subsubsection{Objective Comparisons}
From Fig.\ref{Fig4}, it can be observed that, at high bit-rate in relative, Our(J) with more iterations has more SSIM, PSNR gains than Our(J) with less iteration, no matter what kind of re-sampling is taken among full-resolution re-sampling and low-resolution re-sampling. In the meanwhile, Our(J) should be have less iteration during training at the case of low bit-rate. Our(J)-L has better performance than Our(J)-F on objective measurements below certain low bit-rate about 0.4 bpp, since image can't be well restored from full-resolution re-sampled vectors, when each pixel has very little bits to be assigned, that is to say, each pixel's quality is very low.

In our experiments, these low-resolution re-sampled vectors are compressed with QF sets 2, 6, 10, 20, 30, 40, 50, 60. Meanwhile, full-resolution sampled images are compressed with QF sets 2, 6, 10, 20, 30, 50, 60. Based on the performance on validation dataset of General-100, the final results of Our(J) are performed as follows: Our(J)-L1 (QF=2, 6, 10, 20, 30, 40) and Our(J)-H3 (QF=10, 20, 30, 50, 60), which are displayed in Fig.\ref{Fig5}. At the same time, we also give the objective quality comparisons and iterative number's effects on the performance using our JPEG-compliant image compression with Algorithm-1, when performing on testing dataset mentioned above, as shown in Fig.\ref{Fig6}.

As displayed in Fig.\ref{Fig5},  Our(J) has the best performance on all the testing datasets at both low and high bit-rates in terms of SSIM, as compared to JPEG, JPEG2000, and several combinatorial methods: DicTV \cite{r15}, Foi's \cite{r10}, CONCOLOR \cite{r17}, ARCNN\cite{r19}. For objective measurements of PSNR, Our(J) gets more gains than DicTV \cite{r15}, Foi's \cite{r10}, CONCOLOR \cite{r17}, ARCNN \cite{r19} in most cases, when these methods are compared with JPEG. Among these combinatorial methods, CONCOLOR has better objective measurements than DicTV, Foi's and ARCNN, while DicTV has the worst performance.

When testing on Set5 and Set7, Our(J) can compete with or even better than JPEG2000 in the aspects of PSNR. But Our(J) has lower value than JPEG2000 when testing on Set14, and LIVE1. Since our compressive loss explicitly uses the DSSIM loss for RSN network, image's structures in our re-sampled vectors have protected, which leads to have better structural preservation for Our(J), as compared to others.

From Fig.\ref{Fig5}, it can be also clearly seen that Our(J) has more SSIM and PSNR gains in the whole range of bit-rate against Jiang's \cite{r25}, which can well prove that our algorithm is better than the one of \cite{r25}. It also indicates that our virtual codec network can effectively bridge the gap between RSN network and IDN network. Note than Jiang's \cite{r25} only considers image compression at low bit-rate, but our method can satisfy client's different requirements.

\subsubsection{Visual Comparisons}

Before image decoding reconstruction comparisons between different compression methods, our re-sampled vector by RSN's down-sampling is first compared with Jiang's compact representation \cite{r25}, as shown in Fig.\ref{Fig8} (b1-b3, d1-d3), from which we can see that our re-sampled vector's image can more accurately high-light image's key features. Apart from the down-sampling comparison, we also compare our full-resolution re-sampling with Jiang's down-sampling compact representation at high bit-rate, as displayed in Fig.\ref{Fig9}. Meanwhile, we also compare our compressed re-sampled vector with Jiang's compressed compact representation in Fig.\ref{Fig8} (c1-c3, e1-e3) and Fig.\ref{Fig9} (c1-c3, e1-e3). From these comparison, it can be also concluded that down-sampling compact representation can't burden image's more information, when more bit-rate beyond certain value is assigned for the compression of compact representation. This further turns out that our full-resolution re-sampling is meaningful and efficient to satisfy the scenario of image compression at high bit-rate.

From the (f1-m1, f2-m2, f3-m3) of Fig.\ref{Fig8} and Fig.\ref{Fig9}, it can be noticed that Our(J) preserves image's more structural details than other methods: JPEG, JPEG2000, DicTV \cite{r15}, Foi's \cite{r10}, CONCOLOR \cite{r17}, and ARCNN\cite{r19}. Meanwhile, our method is free of coding's blocking and ringing artifacts than JPEG and JPEG2000. Among these combinatorial approaches \cite{r15,r10,r17,r19}, CONCOLOR and ARCNN have better visual quality than others, while ARCNN's decoded image has a little higher visual quality than CONCOLOR's.

\subsection{Quantitative and qualitative evaluation between DNNC framework and SCIC framework as well as standard codecs}

\subsubsection{Objective Comparisons}
To further demonstrate the effectiveness of our DNNC framework, we compare DNNC framework's results with SCIC framework as well as standard codec. From Fig.\ref{Fig10}, it can be found that Our(D)'s SSIM measurements are better than JPEG for all the testing dataset, especially at low bit-rate, and Our(J) has better performance than JPEG2000. Our(D) even can compete against JPEG2000, when testing data-set of Set5 and Set7. However, Our(D)'s SSIM measurements testing on Set14 and LIVE1 is lower than Our(J) and JPEG2000. Meanwhile, coding efficiency of Our(D) is better than JPEG in term of PSNR at low bit-rate, but is lower than JPEG at high bit-rate.  Besides, Our(J) can compete with JPEG2000 when testing on Set5 and Set7 for PSNR, but JPEG2000's PSNR are large than Our(J)'s on Set14, and LIVE1.
\subsubsection{Visual Comparisons}
The visual comparisons are displayed in Fig.\ref{Fig11}, from which we can see that both Our(D) and Our(J) are free of blocking artifacts and ringing artifacts around discontinuities, when compared to standard codec such as JPEG and JPEG2000. From this figure, we can also observe that JPEG2000 has better visual quality than JPEG, but they are less than Our(D) and Our(J). Although both of Our(D) and Our(J) compress image with high quality, they have different structural and textual preservation at the boundary of image. Beside, images compressed by Our(J) have more smoothness than the ones of Our(D).

\section{Conclusion}
In this paper, an image re-sampling compression method is proposed to efficiently compress image. We generalize this method for SCIC framework and DNNC framework. Due to the intractable problem of learning the whole framework directly, so we decompose this challenging optimization problem into three sub-problems learning. Furthermore, because our coding frameworks are built on auto-encoder architecture, whose output reproduces the input, we can initialize our networks from pre-trained auto-encoder networks. Experimental results have shown that the proposed method is versatile and effective.




\bibliographystyle{IEEEtran}
\bibliography{IEEEfull,VCODEC}
\end{document}